\documentclass[aps,prl,twocolumn,groupedaddress, showpacs]{revtex4}
\usepackage{graphicx}
\usepackage{amssymb}
\usepackage{color}

\newcommand{\hidden}[1]{}

\begin{document}


\title{GHz Spin Noise Spectroscopy in $n$-Doped Bulk GaAs}

\author{Georg M. M{\"u}ller}
\email[Electronic mail: ]{mueller@nano.uni-hannover.de}
\author{Michael R{\"o}mer}

\author{Jens H{\"u}bner}
\author{Michael Oestreich}
\affiliation{Institut f{\"u}r Festk{\"o}rperphysik, Leibniz
Universit{\"a}t Hannover, Appelstra{\ss}e 2, D-30167 Hannover,
Germany}


\date{\today}

\begin{abstract}
We advance spin noise spectroscopy to an ultrafast tool to resolve
 high frequency spin dynamics in semiconductors. The optical
non-demolition experiment reveals the genuine origin of the
inhomogeneous spin dephasing in $n$-doped GaAs wafers at densities
at the metal-to-insulator transition. The measurements prove in conjunction with
depth resolved spin noise measurements that the
broadening of the spin dephasing rate does not result from thermal
fluctuations or spin-phonon interaction, as suggested previously, but from surface
electron depletion.
\end{abstract}


\maketitle


Spin noise spectroscopy (SNS) has developed into a universal tool
to study the spin dynamics in quantum optics and solid state
physics. The technique utilizes the ever present spin fluctuations
in electronic spin ensembles at thermal equilibrium and probes the
spin fluctuation dynamics by absorption-free Faraday rotation.
First introduced by Aleksandrov and Zapassky in a gas of sodium
atoms \cite{aleksandrov:jetp:54:64:1981},
 the technique has been successfully transferred by
Oestreich \textit{et al.} \cite{oestreich:prl:95:216603:2005} to
semiconductor physics. This transfer is of special interest in the
context of semiconductor spintronics since SNS is - in contrast to
the manifold Hanle \cite{parsons:prl:23:1152:1969}, pump-probe Faraday rotation \cite{baumberg:prb:50:7689:1994,kikkawa:prL:80:43131998}, and time- and
polarization resolved photoluminescence experiments \cite{heberle:prl:72:3887:1994} - a weakly
perturbing technique that minimizes optical transitions from the
valence to the conduction band \cite{romer:rsi:78:103903:2007}.
SNS thereby avoids (a) heating of the carrier system which changes
the spin interactions and the spin dephasing rates, (b) creation
of free electrons which inevitably interact and change the spin
dynamics of localized electrons, and (c) the creation of holes
which drives the electron spin dephasing  by the Bir
Aronov Pikus mechanism, i.e., in several cases only SNS yields the
intrinsic, unperturbed spin dephasing rates \cite{muller:prl:101:206601:2008, roemer2009}.
 At the same time,
SNS uniquely allows three dimensional spatial mapping of the spin dynamics \cite{roemer:apl:94:112105:2009}
and promises even single spin detection without destroying the
spin quantum state. However, up to now these advantages could not
be used to its full extend, since the temporal dynamics detected
by SNS has been technically limited by the
available bandwidth of the detection setup. Due to the in many cases low ratio of spin noise power to background noise, efficient data averaging via analog-to-digital conversion and subsequent FFT as well as laser noise suppression via balanced photodetection is necessary. Therefore, highly sensitive SNS is technically limited by the bandwidth of commercially available digitizers and low noise balanced photoreceivers to frequencies of about 1~GHz \cite{crooker:unpublished}.

In this letter, we demonstrate in $n$-doped GaAs -- probably the
most prominent model system in semiconductor spintronics -- that
ultrafast, highly sensitive SNS \cite{starosielec:apl:93:051116:2008} becomes feasible by sampling the
temporal dynamics of the electron spin with the laser repetition
rate of a picosecond Ti:Sapphire laser oscillator. In the first
part, we present the technique and argue that ultrafast SNS
is applicable up to THz frequencies. In the second part, we
combine ultrafast and depth resolved SNS measurements
\cite{roemer:apl:94:112105:2009} to explore the physical origin of
the inhomogeneous spin dephasing in $n$-doped GaAs close to the
metal-to-insulator transition (MIT) at high magnetic fields and
study the field dependence of the electron Land\'e $g$-factor.



\begin{figure}[tbh]
    \centering
        \includegraphics[width=\columnwidth]{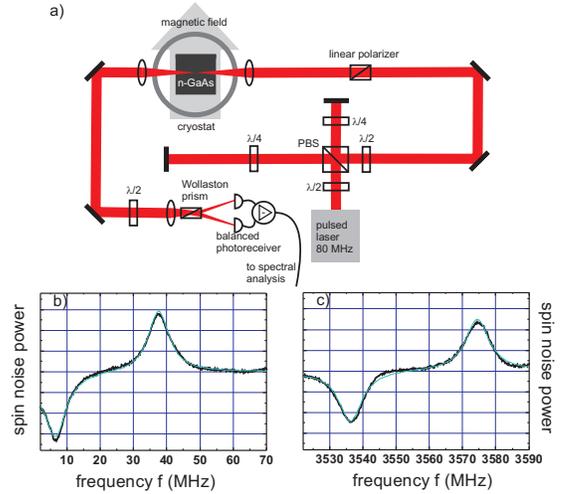}
    \caption{(Color online) (a) Experimental setup for
     GHz SNS. The repetition rate of the
    pulsed laser is doubled via a
    Michelson interferometer like setup. (b, c) Spin noise
    spectra acquired from sample A at 25~K; the peaks
    correspond to 6.0 and 1.0~mT (b) and to  594.7
    and 588.3~mT (c), respectively. The spectra are corrected
    for frequency dependent amplification via division by the
    amplification curve. The light solid lines are Lorentzian line fits to the data.}
    \label{fig:setup}
\end{figure}

Figure~\ref{fig:setup}(a) schematically depicts  the high
frequency SNS experimental setup. The repetition rate of an
actively mode-locked ps-Ti:Sapphire laser ($f_{\mathrm{rep}}
=80$~MHz) is doubled to $f_{\mathrm{rep}}'=2f_{\mathrm{rep}}$ via
a Michelson inter\-fero\-meter like setup with different arm lengths.
The linear polarized ps laser pulses  are transmitted through a
$n$-doped GaAs sample where the wavelength of the laser is tuned
below the GaAs band gap to avoid absorption. The mean spin
polarization of the electron ensemble in the GaAs is zero at
thermal equilibrium but the root mean square deviation is unequal
to zero and fluctuates with the transverse spin lifetime $T_2^\ast$.
The spin fluctuations are mapped via  Faraday rotation onto the
direction of the linear polarization of the laser light which is
measured by a polarization bridge. The resulting electrical signal
from the balanced photoreceiver is amplified by a low noise
amplifier, frequency filtered by a 70~MHz low pass filter,
seamlessly digitized in the time domain (sampling rate 180~MHz)
and spectrally analyzed in real time via fast Fourier transform. 
An
external magnetic field $B$ in Voigt geometry modulates the spin
fluctuations with the Larmor precession frequency
$f_{\mathrm{L}}=g^\ast\mu_{\mathrm{B}}B/h$, where $g^\ast$ is the
electron Land\'e $g$-factor.
By sampling the spin dynamics with the stroboscopic laser light, the spin noise spectrum $S(f)$  centered  around $f_{\mathrm{L}}$  evolves into a sum of noise peaks $\sum_n{S(f-nf_{\mathrm{rep}}')}$. The limit of the summation index $n$ is given by the ratio of the inverse pulse length to the repetition rate.
Due to the doubling of the original laser repetition rate, only one summand of
this  sum falls into the detection window.
In principle, utilizing a  GHz digitizer card and a fs laser system with GHz repetition rate, both commercially available, increases the bandwidth of the detection and the detectable line width to a GHz and the detectable Larmor frequencies to the THz regime.
 We want to point out that this sampling with pulsed laser light  is not a nonlinear
process like  electrical frequency mixing and thereby does not
introduce any additional noise.

The two investigated samples $A$ and $B$ are commercial Si-doped
GaAs wafers with doping densities right at the metal-to-insulator transition (MIT)
($n_{\mathrm{d}}^{\mathrm{A}}=1.8\cdot 10^{16}\,\mathrm{cm}^{-3}$)
and about 4 times above the MIT
($n_{\mathrm{d}}^{\mathrm{B}}=8.8\cdot 10^{16}\,\mathrm{cm}^{-3}$)
and thicknesses of 340~$\mu$m and 370~$\mu$m, respectively. Both
samples are anti-reflection coated to increase the light
transmittance. The samples are mounted in a He gas flow cryostat
with a superconducting split coil magnet in Voigt geometry. The
magnetic field is carefully calibrated by means of a Hall probe
and the magnet system is always degaussed before starting a new
series of measurement at lower magnetic fields to avoid magnetic
hysteresis. The laser wavelength is set to 840 (845) nm for
measurements concerning sample A (B) if not stated otherwise.

Figure~\ref{fig:setup}(b) and (c) depict two typical spin noise
spectra of sample A at 25~K. Electronic and optical shot noise is
eliminated in these spectra by subtracting in each case two noise
spectra at slightly different magnetic fields which results in a
spin noise spectrum with one positive and one negative spin noise
peak. The width and position of these two peaks are determined by
fitting a sum of Lorentz lines to the experimental spectra. The
resulting full width at half maximum of each Lorentz peak $\Delta
f_{\mathrm{FWHM}}$ and the peak position yield the spin dephasing
rate $\Gamma_{\mathrm{s}}=1/T_{2}^\ast=\pi \Delta f_{\mathrm{FWHM}}$ and $g^\ast$,
respectively. The acquisition times of both spectra are less than
20 minutes. The excellent signal to noise ratio of the high
frequency spectrum (Fig.~\ref{fig:setup}(c)) matches the signal to
noise ratio of the low frequency spectrum
(Fig.~\ref{fig:setup}(b)) proving that the ultrafast SNS
comes along without loss of sensitivity.

\begin{figure}
    \centering
        \includegraphics[width=\columnwidth]{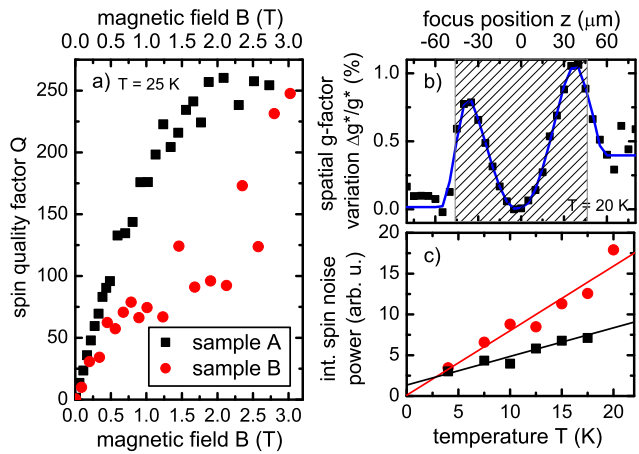}
    \caption{(Color online) (a) Spin quality factor
   as a function of magnetic field. In sample B, the spin noise power drops for high fields which we attribute to Pauli blocking of spin noise due to filling of the Landau levels. (b) High aperture SNS
    measurement: Shift of the electron $g$-factor in dependence of the
    position of the laser focus relative to the sample (indicated by the shaded area).
    The line is a guide to the eyes only. The sample appears compressed by the refractive index of GaAs. Note that the average g-factor variation with the laser focus outside the sample is lower than  the average value with the focus inside. This indicates that the spin noise in the sample center  contributes stronger to the GHz  spin noise measurements (large Rayleigh range) than the spin noise from the sample surfaces. (c) Integrated spin
    noise power as a function of temperature (corrected for different sample thickness; laser wavelength:
    840~nm, magnetic field: 6.75~mT). The lines are linear fits.}
    \label{fig:rates}
\end{figure}

Next, we employ high frequency SNS and measure the transverse
magnetic field dependence of $\Gamma_{\mathrm{s}}$ and $g^\ast$ in bulk
$n$-GaAs.  Optical pumping
techniques clearly influence both $\Gamma_{\mathrm{s}}$
\cite{putikka:prb:70:113201:2004} and $g^\ast$
\cite{lai:apl:91:062110:2007}. We focus on the doping regime close to the MIT where an
intricate interplay between localized and free electrons leads to
a maximum in the low temperature spin lifetime
\cite{dzhioev:prb:66:245204:2002}.  Many theoretical and experimental
groups have already investigated this doping regime but the
physical origin of the experimentally observed homogeneous or
inhomogeneous broadening of $\Gamma_{\mathrm{s}}$ with increasing transverse
magnetic fields was up to now unsolved. Kikkawa and Awschalom for
example measured via resonant spin amplification a plateau of the
spin quality factor $Q=g^\ast\mu_{\mathrm{B}}BT^\ast_2/h$ at around
80 for a temperature of 5~K \cite{kikkawa:prL:80:43131998}. They
attributed this maximal $Q$-factor -- which is a measure of the
reduction of the spin lifetime in a transverse magnetic field --
to the thermal electron distribution and the energy dependence of
$g^\ast$ yielding an inhomogeneous broadening. Puttika and Joynt gave a different explanation for the
same experimental data and suggested a spin-phonon mechanism,
which effectively leads to a homogeneous broadening of the spin dephasing rate for
localized electrons \cite{putikka:prb:70:113201:2004}. 
Figure~\ref{fig:rates}(a)
shows the  $Q$-factor  of a similar sample (sample A) measured via ultrafast SNS at a lattice temperature
of 25~K \footnote{The experimental values given in this letter
correspond to the most symmetric measured spin noise peaks and,
hence, to the longest measured spin lifetimes, since some traces
show an asymmetric behavior from unknown origin. However, no
systematic dependence on the angle between sample surface or
magnetic field direction can be reported.}. We measure a $Q$-factor
that is despite a five times higher
temperature about three times larger than the value reported in
Ref.~\cite{kikkawa:prL:80:43131998}. This finding directly
disproves 
that the  $Q$ measured by Kikkawa and Awschalom is limited by a $g$-factor spread resulting
from a thermally broadened electron energy distribution. Such a
thermally broadened $g$-factor spread is negligible on these
timescales due to a motional narrowing type averaging in the electron energy
\protect\cite{zutic:rmp:76:32:2004}, which, additionally, leads to a quadratic dependence of the broadening on the magnetic field and, hence, would not result in the formation of a $Q$-factor plateau. The ultrafast SNS
measurements also discourage the theory of thermally activated
lattice vibrations, since the
line shape of the SNS spectrum shows a clear crossover from a
homogeneously broadened Lorentzian peak to an inhomogeneously
broadened Gaussian type peak \footnote{At moderate fields, a physical correct fit
function would base upon Voigt profiles which have too many free
parameters for practical purposes. Nevertheless, fitting with
Gaussian profiles yields very similar peak widths as fitting
Lorentzian profiles. At small magnetic fields and temperatures below 50~K, the temperature dependence of $\Gamma_{\mathrm{s}}$ scales with the temperature dependence of the conductivity \cite{roemer2009} suggesting a spin dephasing mechanism that depends on motional narrowing and should therefore be less efficient at higher $B$ \cite{kavokin:sst:23:114009:2008}.}.

For further investigations of the origin of the inhomogeneous
broadening, we use a high aperture cw spin noise setup (see
Ref.~\cite{roemer:apl:94:112105:2009} for details) in order to
check for spatial electron density variations within sample A that could,
e.g., come from the Czochralski growth method. The unique
spatial resolution strength of SNS originates from the fact that
the relative noise power becomes large for small electron
ensembles so that most of the SNS signal is acquired within the
Rayleigh range of the laser focus. The depth resolved measurements are
carried out in a micro-cryostat at a lattice temperature of 20~K,
a laser wavelength of 830 nm, a Rayleigh range in vacuum of
$11\,\mathrm{\mu m}$ and accordingly a depth resolution in the
sample of about $80\,\mathrm{\mu m}$, and a magnetic field of
about 9~mT. Figure~\ref{fig:rates}(b) shows the measured variation
of the $g$-factor $\Delta g^\ast/g^\ast$ versus focus position.
The focus position is swept in this experiment from negative to
positive $z$-position where the $z$-axis is chosen along the direction of light propagation and we define $\Delta g^\ast /g^\ast=0$ at
$z=0\,\mathrm{\mu m}$. The measurements reveal that the absolute
value of $g^\ast$ has a pronounced maximum close to the front and
the back surface of the sample. Such a maximum of $|g^\ast|$ is
directly linked at this doping concentration via the energy
dependence of the $g$-factor to a minimum in electron density \cite{yang:prb:47:6807:1993}. The
width of the two measured peaks is equal to the depth resolution
of the laser which indicates that the spatial variation of
$g^\ast$ is smaller than the depth resolution of the experiment.
Conductivity measurements in sample A evince hopping behavior \cite{roemer2009}. This $g$-factor variation is not observed at higher temperatures where the conductivity significantly increases and most electrons reside in the conduction band.

We model the spatial variation of the measured $g$-factor by a
simple phenomenological model assuming partial electron depletion
at the surface, which results in a decrease of the $g$-factor, the
spin lifetime, and the noisepower at the sample surface, and get
qualitative agreement with the measured SNS profile.Other reasons for the observed spatial g-factor broadening could involve stress in the sample or specifics of the sample growth we are not aware of.
A quantitative description is difficult due to the intricate
interplay between localized and free electrons, Fermi level
pinning, the influence of hopping and diffusion on the
depletion zone, and the electric field induced change of $g^\ast$,
the band gap, and the interband transition probability  \footnote{In
our  simple model, we have to assume a width of the depletion
zone of some ten microns in order to mimic the measured results. A
more sophisticated model would probably yield a different width. Nevertheless, we are not aware of any theoretical description of depletion zones at the MIT. Additionally, spin diffusion and electron hopping also lead to smoothing of the measured $g$-factor profile which is not accounted for in our modelling.}.
Nevertheless, the spatially resolved $g^\ast$-variation directly
entails the inhomogeneous broadening measured by ultrafast SNS where a Rayleigh range much larger than the sample thickness is employed.
The much smaller
$Q$-factor reported in Ref.~\cite{kikkawa:prL:80:43131998} most likely originates from  the lower conductivity  at 5~K and the slightly lower doping concentration in accordance with our reasoning above.

\begin{figure}
    \centering
        \includegraphics[width=\columnwidth]{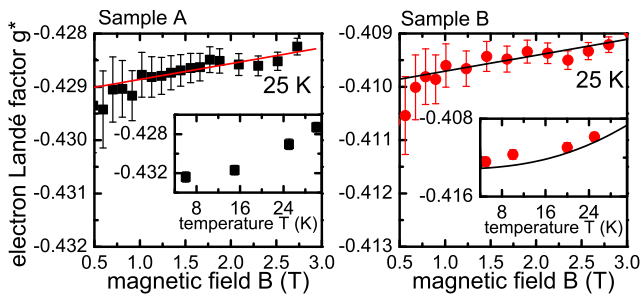}
    \caption{(Color online) Electron Land\'e $g$-factor as a function
    of magnetic field  for  sample A and B at 25~K and as a
    function of the sample temperature at a magnetic field around 850~mT.
The error bars of the magnetic field dependence include the
inaccuracy of the magnetic field and the repetition rate
$f_{\mathrm{rep}}'$ and the fitting errors.
}
    \label{fig:gfactor}
\end{figure}

Figure~\ref{fig:rates}(b) shows as a second feature a spatial
asymmetry of the SNS signal, i.e., the right peak is higher than
the left peak and $\Delta g^\ast /g^\ast$ is larger if the laser
focus is behind the sample ($z>46~\mu$m) compared to the laser
focus in front of the sample ($z<-46~\mu$m). We have carefully
checked that this increase does not result from a doping gradient
in the sample but from the finite measurement time, i.e., $\Delta
g^\ast /g^\ast$ increases with laboratory time and saturates after about
one hour. To account for this effect, experimental data presented
in Fig.~\ref{fig:rates}(a) and \ref{fig:gfactor} are acquired
after the saturation of this effect. 
We attribute  this behavior to the existence of deep centers acting as an electron trap, like the EL2 deep center, 
which has a metastable state with a large lattice
relaxation and  is excited by below band gap excitation
\cite{mitonneau:ssc:30:157:1979,martin:apl:39:747:1981,
vincent:jap:53:3643:1982}. Such deep centers are the origin
of various persistent photo effects in liquid phase grown GaAs.
We exclude nuclear effects as origin since the
spatial measurements were carried out at very low magnetic fields
and since the integrated spin noise power also changes with time.
Within the measurement error, the spin lifetime is not affected by this effect. 

Figure~\ref{fig:rates}(a) also shows the $Q$-factor for sample~B
at 25~K. Within the examined field range the $Q$-factor does not level off as in sample~A.   The doping density of sample~B of $n_{\mathrm{d}}^{\mathrm{B}}=8.8\cdot
10^{16}\,\mathrm{cm}^{-3}$ is well above the MIT and the electrons
should therefore be delocalized. In order to directly prove the
delocalized nature of the electrons in sample~B, the integrated
spin noise power is plotted in Fig.~\ref{fig:rates}(c) for both
samples as a function of temperature. The extrapolation to zero
temperature yields for sample~A a finite spin noise power proving partial
localization of electrons. The same extrapolation yields for
sample~B within the error bars zero spin noise power which is
consistent with delocalized electrons and Pauli blocking of the
spin noise \footnote{Crooker and co-workers reported for a lower
doping density of $7.1\cdot 10^{16}\,\mathrm{cm^{-3}}$ still a
finite fraction of spin noise power at extrapolation to zero
temperature \cite{crooker:prb:79:035208:2009}.}.  To our knowledge, there is no effective mechanism of inhomogeneous broadening of the spin dephasing rate for free electrons in this sample system.  Hence, a strictly monotonic increase of the $Q$-factor with the magnetic field is expected in sample B.

Last, we study the magnetic field and temperature dependence of
$g^\ast$. To our knowledge, similar investigations were only
carried out for undoped or very low $n$-doped bulk GaAs samples
\cite{oestreich:prb:53:7911:1996, seck:prb:56:7422:1997}.
Figure~\ref{fig:gfactor} shows the magnetic field dependence of
$g^\ast$ for sample A and B at 25~K. A linear fit to the magnetic
field dependence of $g^\ast$ between 0.5~T and 3~T yields a
very small magnetic field dependence of
$g^\ast_{\mathrm{A}}=-0.4292(2)+0.0003(1)\, \mathrm{T}^{-1}\cdot B$ for
sample A and
$g^\ast_{\mathrm{B}}=-0.4100(2)+0.0003(1)\,\mathrm{T}^{-1}\cdot B$ for
sample B at $T=25\,\mathrm{K}$. The gradient $dg^\ast/dB$ is by more than
one order of magnitude smaller than for free electrons in undoped
GaAs \cite{oestreich:prb:53:7911:1996} or for donor-bound
electrons at higher magnetic fields \cite{seck:prb:56:7422:1997},
where $g^\ast(B)$ increases in both cases by
$0.005\,\mathrm{T}^{-1}$ due to the energy shift of the lowest Landau level which is occupied by all electrons. At the  doping densities examined in this work, either Landau quantization is suppressed because of momentum scattering or several Landau cylinders are occupied since the cyclotron energy is rather low compared to the Fermi energy. 
 The difference of $g^\ast$ between the two samples extrapolated
to $B=0$ is about $0.02$ which equates to a difference
in Fermi energy of about 3~meV if we adapt the
experimentally and theoretically reported energy dependence of
$g^\ast$ in GaAs of 6.3~eV$^{-1}$ (see Ref.~\cite{yang:prb:47:6807:1993} and references therein). This difference in the Fermi energy is
reasonable, i.e., the calculated difference in the Fermi level is smaller assuming
only an impurity band \cite{shklovskiiefros1984} and larger assuming only free
electrons in bulk GaAs. The same energy dependence of
6.3~eV$^{-1}$ explains quantitatively the absolute value and the
temperature dependence of $g^\ast$ for sample~B. The solid line in
the right inset of Fig.~\ref{fig:gfactor} depicts $g^\ast (T)$
calculated by
\[
g^\ast(T)=\frac{\int_0^\infty DOS(E) f(E,T) (1-f(E,T)) g^\ast(E)
dE}{\int_0^\infty DOS(E) f(E,T) (1-f(E,T)) dE},
\]
where $DOS(E)$ is the three dimensional density of states,
$f(E,T)$ is the Fermi distribution, $g^\ast(E)=g^\ast_0 +
6.3\,\mathrm{eV}^{-1}\cdot E$, and $g^\ast_0$ is the electron Land\'e
$g$-factor at the conduction band minimum at $T=0$ which is set to
$g^\ast_0 = -0.481$ in agreement within the error bars of
the high precession measurements of
Ref.~\protect\cite{hubner:prb:79:193307:2009}. The three
dimensional density of states is a good approximation for highly
doped samples. For sample~A, the measured temperature dependence
of $g^\ast$ is slightly larger than in sample~B and is more
difficult to calculate since donors at the MIT form an impurity
band and ionization of electrons into the conduction band has to
be included.

In conclusion, we have overcome the major limitation of
conventional spin noise spectroscopy and demonstrated ultrafast
SNS without any loss of sensitivity. We have applied the technique
to $n$-doped GaAs at and above the metal-to-insulator transition
and proved in conjunction with depth-resolved SNS that in the
sample at the MIT surface depletion changes the $g$-factor locally resulting in inhomogeneous spin dephasing at high magnetic fields.  This mechanism is of extrinsic nature and samples at the MIT with a spatially flat Fermi level should yield significantly higher $Q$-factors.
Measurements on the temperature dependence of the spin noise power
verify the (de)localization of the electrons in the sample at (above) the MIT and ultrafast spin noise
measurements yield the magnetic field and temperature dependence
of $g^\ast$ for thermalized electrons in the impurity and in the conduction band. Most importantly,
we want to point out that ultrafast SNS is not limited to
semiconductor spintronics but promises access to ultrafast spin
dynamics, e.g., in atomic ensembles,  magnetically ordered
systems, like ferrimagnetic garnets where recently the Bose-Einstein condensation of magnons was discovered \cite{demokritov:nat:443:430:2006}, or in superconductors at the phase transition \cite{Prozorov2008}.

This work was supported by the German Science Foundation (DFG
priority program 1285 `Semiconductor Spintronics'), the Federal
Ministry for Education and Research (BMBF NanoQUIT), and Centre
for Quantum Engineering and Space-Time Research in Hannover
(QUEST).
G.M.M. acknowledges support from the Evangelisches Studienwerk.\\

\end{document}